\documentclass[3p,twocolumn]{elsarticle}
\usepackage{lineno,hyperref}
\usepackage{xcolor}
\usepackage{physics}
\usepackage{txfonts}
\usepackage{xspace}
\usepackage{graphicx}
\newcommand{\Uttf}{\mbox{$^{235}\textrm{U}$}\xspace}
\newcommand{\Utte}{\mbox{$^{238}\textrm{U}$}\xspace}
\newcommand{\Puttn}{\mbox{$^{239}\textrm{Pu}$}\xspace}
\newcommand{\Putf}{\mbox{$^{240}\textrm{Pu}$}\xspace}
\newcommand{\UtOe}{\mbox{$\textrm{U}_{3}\textrm{O}_{8}$}\xspace}
\newcommand{\PuOt}{\mbox{$\textrm{Pu}\textrm{O}_{2}$}\xspace}
\newcommand{\mycomment}[1]{}
\modulolinenumbers[5]

\begin{document}

\begin{frontmatter}

\title{Machine Learning technique for isotopic determination of radioisotopes using HPGe $\mathrm{\gammaup}$-ray spectra \tnoteref{mytitlenote}}

\author{Ajeeta Khatiwada}
\cortext[mycorrespondingauthor]{Corresponding Author}
\ead{ajeeta@lanl.gov}
\author{Marc Klasky}
\author{Marcie Lombardi}
\author{Jason Matheny}
\author{Arvind Mohan}
\address{Los Alamos National Laboratory, Los Alamos, NM 87545, USA}

\begin{abstract}
    $\mathrm{\gammaup}$-ray spectroscopy is a quantitative, non-destructive technique that may be utilized for the identification and quantitative isotopic estimation of radionuclides. Traditional methods of isotopic determination have various challenges that contribute to statistical and systematic uncertainties in the estimated isotopics. Furthermore, these methods typically require numerous pre-processing steps, and have only been rigorously tested in laboratory settings with  limited shielding.  In this work, we examine the application of a number of machine learning based regression algorithms as alternatives to conventional approaches for analyzing $\mathrm{\gammaup}$-ray spectroscopy data in the Emergency Response arena. This approach not only eliminates many steps in the analysis procedure, and therefore offers potential to reduce this source of systematic uncertainty, but is also shown to offer comparable performance to conventional approaches in the Emergency Response Application.
\end{abstract}


\end{frontmatter}

%\linenumbers

\section{Introduction}
\label{sec:intro}

%Radioactive sources produce $\mathrm{\gammaup}$-ray and neutrons at different energies and intensities. The emitted $\mathrm{\gammaup}$-rays interact with the detector material of a $\mathrm{\gammaup}$-ray spectrometer to produce gamma spectrum. The photo-peaks at discrete energies, referred to as lines, are utilized in $\mathrm{\gammaup}$-ray spectroscopy technique to identify the radioactive isotopes and their fraction in a sample.\par

The identification and quantitative determination of the isotopic content of samples/objects potentially containing uranium and/or plutonium  is of paramount importance to the nuclear materials safeguards, arms control verification,  nuclear security, Emergency Response (ER), as well in nuclear remediation arenas~\cite{walton1974measurements, gunninkmgau, gunnink1990mga, wei2021application, peyvandi2018determination, connor2016airborne, hosoda2016environmental, korob2006simple, apostol2016isotopic, aitkenhead2012use}. Conventional methods for determining the isotopics/enrichment using $\mathrm{\gammaup}$-ray spectroscopy require many time consuming steps
\begin{enumerate}
    \item photo-peak identification,
    \item background and continuum subtraction,
    \item feature extraction,
    \item estimation of the relative efficiency curve, and
    \item matching of the extracted features with those of known nuclides to estimate the fraction of isotopes~\cite{rawool2010steps}.
\end{enumerate}

 In many of these application areas, it is imperative to rapidly determine the isotopic fractions using remote detection techniques. These constraints necessitate the use of non-destructive assay methods (NDA) and accompanying automated algorithms to perform quantitative analysis. In some applications, details regarding the physical arrangement of the nuclear materials cannot be revealed due to security concerns, e.g. in treaty verification activities, or are unknown e.g. in nuclear security and ER activities in which the shielding and other aspects of the physical configuration are unknown. In this work, we examine the ability of numerous machine learning (ML) techniques to address the automated identification and quantification of  uranium and plutonium isotopics for ER applications. 
\section{Organization of Paper}
In this work, we investigate the application of a variety of machine learning algorithms to perform uranium and plutonium isotopic estimation for Emergency Response applications. Before discussing the ML algorithms utilized in these investigations, we present a review of both the traditional as well as the ML methods to perform quantitative isotopic identification in Section~\ref{sec:background}.  The machine algorithms utilized in this investigation are presented in Section~\ref{sec:ML}.  In Section~\ref{sec:datagen}, the generation of ML training data is discussed along with an investigation of the accuracy of these simulations to emulate experimental data. Details of the pre-processing of the spectral data including background, continuum subtraction, and feature extraction are given in Section~\ref{sec:pre}. 
%Details on the training procedures and hyper-parameter selection for the ML algorithms are provided in Section ~\ref{sec:ML}.
ML results using simulations are presented in Section~\ref{sec:sim}.  Hyper-parameter investigations are presented in Section~\ref{sec:Hyper}. Investigations using experimental data and discussions of the results are presented in Section~\ref{sec:Exp}.  Lastly, summary and conclusions are provided in  Section~\ref{sec:Conclusions}.\par

\section{Background}
\label{sec:background}

\subsection{Traditional Methods}

Starting in the early 1970s, researchers developed several approaches to perform quantitative NDA spectroscopic analysis for both uranium and plutonium isotopics~\cite{parker1974plutonium}.
Today there are three general variations of the NDA method that have been utilized to infer the isotopic content of \Uttf. The first method, currently utilized by the International Atomic Energy Agency (IAEA), is based on the measurement of the 186~keV line of \Uttf\ in the spectra obtained using either germanium or sodium iodide spectrometer systems~\cite{reilly1970progress}, and requires a calibration with a known enrichment standard. Provided that the sample measured is similar to the reference i.e. has the same geometry and thickness and the measurement conditions are constant, the counting rate for the 185.7~keV peak is proportional to the enrichment.  While this approach has been utilized to successfully infer the content of \Uttf, there are several limitations: the samples must satisfy the infinite thickness criterion~\cite{reilly1991passive}, calibrations need to be performed for samples with different containers, and wall thicknesses need to be determined prior to the enrichment measurement~\cite{sampson1989fram}. In practice, this constraint limits the applicability of the the enrichment of an object’s surface to a depth of 0.26~cm and 0.74~cm for uranium metal and \UtOe\ powder, respectively~\cite{reilly1991passive}.
An automated version of this method, called NaIGEM (NaI(Tl) Gamma Enrichment Measurements), is included in the HM-5 instrument used by the IAEA~\cite{mortreau2004determination}. Enrichment measurements of uranium without contaminants using low-resolution detectors can achieve 1\% precision for arbitrary enrichment while contamination by minor uranium isotopes has a biasing effect of 5--10\%~\cite{sprinkle1997low}.

Methods employing multi-peak self calibration were proposed to overcome the drawbacks of the enrichment meter principle. The first variation, Peak Area (PA), utilizes the spectral lines in the range 89 to 120 keV~\cite{gunnink1990mga}. The relative efficiency curves of different uranium isotopes or their daughters are estimated from a limited number of peaks in the spectrum. Sophisticated codes such as MGAU (Multi-Group Analysis for uranium) are  based on this principle. The precision of the estimated efficiency response depends on the the number and intensity of the isotope peaks.  However, these methods still experience performance issues when measuring uranium through thick walled containers~\cite{abousahl1996applicability, gunnink1994mgau, gunninkmgau, morel1978references, morel2000results}.  

To overcome the limitation of the finite thickness of shielding, the relative-efficiency (RE) method was proposed~\cite{parker1974plutonium}. The RE method
computes the uranium enrichment using the relative efficiency
obtained from the peaks expressed in the measured spectra using an energy range from 144 to 1001 keV. Several software packages, including FRAM and  MGA++\footnote{A suite of three software programs (MGA, U235, and MGAHI, a Pu isotopic analysis code that uses the 200 keV -1 MeV energy region) for the analysis of actinide spectra acquired by Ge detectors.}, have implemented this approach~\cite{korob2006simple, sampson2003application, darweesh2019study}.
%MGA++ is a suite of three software programs (MGA, U235, and MGAHI, a Pu isotopic analysis code that uses the 200 keV -1 MeV energy region) for the analysis of actinide spectra acquired by Ge detectors.
Both MGA++ and FRAM may be utilized to perform Pu isotopics analysis using the low-energy $\mathrm{\gammaup}$-ray spectrum, along with higher energy $\mathrm{\gammaup}$-rays~\cite{gunnink1990mga, sampson2003application}.

A comparison of three implementations of the RE method concluded that the performance and applicability with increasing wall thickness at low enrichment grades was in the order PC/FRAM, MGA++, and MGAU.  Therefore, in shielded conditions, it was recommended that PC/FRAM for $\mathrm{\gammaup}$-rays above 200 keV using the coaxial detector spectrum be utilized. Before concluding, it should be noted that the shielding thicknesses that were evaluated are significantly below those that might be encountered in ER scenarios \cite{darweesh2019study}, i.e., the shielding thickness may be significantly greater than those analyzed with the traditional approaches for determining uranium enrichment and or plutonium isotopics.

\subsection{Machine Learning Methods}

The traditional methods utilized to perform NDA of uranium enrichment and plutonium isotopic quantification require numerous pre-processing steps, and also have difficulty in treating environments in which unknown shielding, overlapping peaks, and or thick shielding is present.  These issues, in conjunction with the success in the development and application of machine learning (ML) techniques in the last decade, have motivated the examination of  machine learning techniques to address these shortcomings.  Indeed, the application of ML techniques to address both classification and regression problems in radiation detection, source identification, and quantitative assessment of radionuclides applications have become increasingly popular.

One of the more prominent applications of ML in addressing radioisotopes has been in the detection and identification arena for nuclear safeguards and arms control applications. To that effect, one of the first applications of a neural network to identify radioisotopes was performed by Olmos using a low resolution NaI detector~\cite{olmos1992application}. Additional early work by Yoshida utilized a multilayer perceptron (MLP) network with a HPGe spectra to identify radioisotopes  in samples with mixed radioisotopes~\cite{yoshida2002application}. Kangas also developed a neural network to analyze very low resolution Polyvinyl toluene (PVT) spectra for use in the identification of radioactive materials at international border crossings~\cite{kangas2008use}. More recently, Liang has demonstrated that a Convolutional Neural Network (CNN) algorithm trained using Monte Carlo N-Particle Transport (MCNP)~\cite{Briesmeister1993MCNPAGM} simulations with a NaI detector could, in a low count rate regime, identify radioisotopes that are nominally difficult to identify, eliminating the necessity to perform spectra pre-processing such as background subtraction and spectrum smoothing~\cite{liang2019rapid}. Bobin utilized a Bayesian sequential approach combined with a spiking neural network to enable the real-time processing of signals detected from a mixture of $\mathrm{\gammaup}$-emitting radionuclides in spectroscopic portal systems~\cite{bobin2016real}.
Finally, Sharma et al. implemented machine learning techniques  to reduce false alarm rates when using $\mathrm{\gammaup}$-ray spectrometers for the identification of persons concealing radioactive materials~\cite{sharma2012anomaly}.

Additional investigations have been performed in the application of neural networks for radioisotope identification~\cite{he2018rapid, hague2019comparison, kamuda2019automated, zhang2019identification}.
 In general, these investigations utilized either MLP or CNNs with a number of different methods for feature extraction, including the Discrete Cosine Transform (DCT) and the Karhunen-Lo\`eve Transform (KLT).  A more advanced neural network architecture employing an autoencoder with a low resolution NaI detector was shown to improve anomaly detection relative to traditional techniques~\cite{bilton2021neural}.

Another application of ML is in the area of identification of radioisotopics in environmental samples~\cite{gomez2020status, hata2015application}. Hata investigated the feasibility of using a support vector machine (SVM) to classify uranium waste drums as natural uranium or reprocessed uranium using NaI detectors~\cite{hata2015application}. Wei applied a radial basis neural network algorithm for environmental and treatment evaluation of decommissioned uranium tailing ponds~\cite{wei2021application}. Finally, Chen used a KLT and an artificial neural network in conjunction with NaI~\cite{chen2009nuclide}.

Additional application areas of ML have been investigated including the analysis of complex spectra (fission and activation products). In these applications, it was shown that the application of feed forward neural networks in conjunction with the Singular Value Decomposition (SVD) can significantly improve performance and reduce the required analysis time once the neural networks have been trained~\cite{pilato1999application}.
 
In some applications, the objective is to determine the isotopic content of the radioisotopes. In particular, in the ER application the objective is to determine the uranium enrichment and or plutonium isotopics in objects containing nuclear material. In this scenario,  HPGe detectors are typically utilized, and the geometry of the object containing the nuclear material along with the characteristics of the intervening shielding materials, i.e. material composition and thicknesses of the components containing the nuclear materials, are not known.  Although many investigations have been performed using ML algorithms to determine isotopic content, almost all of these have been conducted in applications related to Nuclear Material Safeguards and other application areas in which either the configuration is known and or the shielding materials are both known and/or relatively thin i.e. less than 1 cm. Notably, Shaban utilized a feed forward neural network to predict uranium enrichment in laboratory size samples~\cite{shaban2019applying}.

Early work by Vigneron demonstrated that HPGe spectroscopic measurements in conjunction with Principle Component Analysis (PCA), to reduce the dimensionality of the spectra, could be successfully utilized to determine the enrichment of laboratory samples using the low energy range 83 to 103 KeV using a MLP~\cite{vigneron1996statistical}.

Ryu investigated the use of a neural network model using low resolution NaI spectra to analyze uranium enrichment, from depleted to low enrichment, from very low radioactivity samples present in small beakers with very short count times~\cite{ryu2021development}.
Elmaghraby also utilized a neural network architecture to determine the uranium isotopics using a HPGe detector on laboratory samples~\cite{elmaghraby2019determination}.

Lastly, Aitkenhead  using simulated data, evaluated the spectra of shielded plutonium
using ANNs to detect the presence or absence of plutonium, estimate \Puttn\ content,
as well as distinguish material age of shielded plutonium~\cite{aitkenhead2012use}.

\section{Machine Learning algorithms }
\label{sec:ML}
While a great deal of work has been performed to investigate the use of machine learning in the areas of radioisotope detection and identification as well in the quantification of radioisotopes, almost all of this work has been conducted under conditions that are not directly relevant to the ER community.  Accordingly, in this work we examine the application of machine learning algorithms (MLP and Convolutional) Neural Networks, Gaussian Processes, Decision Tees and their variants i.e.  Gradient Boosted Decision Trees and Random forest, as well as Nearest-Neighbors to 1)  study if ML based regression algorithms are a reasonable alternative to the conventional methods and 2) to identify a general class of ML algorithms that are robust to achieving the aforementioned goal without excessive fine-tuning of the hyper parameters to enable the determination of the isotopic content of uranium and plutonium under conditions more consistent with ER application.

\subsection{Methods}

Examinations in this paper are performed based on supervised learning of training datasets using regression algorithms that are integrated into the Scikit-learn~\cite{scikit-learn} package in Python as well as ML algorithms available in Mathematica~\cite{wolfram2003mathematica}. The results from Mathematica are labeled with `*' next to the algorithm names in the tables.\par

\subsubsection{Decision Tree}
\label{subsec:Decision Tree Methods}

Decision trees are one of the most commonly used, practical approaches for supervised learning. They can be used to solve both regression and classification tasks.  A decision tree builds regression or classification models in the form of a tree structure. They break down a dataset into smaller and smaller subsets while at the same time an associated decision tree is incrementally developed. The final result is a tree with decision nodes and leaf %nodes~\cite{praagman1985classification}.
nodes~\cite{breiman1984classification}.
Each tree is composed of nodes, which are chosen by looking for the optimum split of the features. The split of features is determined utilizing an impurity measure. For regression trees, two common impurity measures are least squares and least absolute deviations.  In the former, the method is similar to minimizing least squares in a linear model. The splits are chosen to minimize the residual sum of squares between the observation and the mean in each node.  In the latter method,
a minimization of the mean absolute deviation from the median within a node is performed. 

Two popular techniques to improve the robustness of a decision tree are ensemble methods such as Random Forest methods and Boosting methods. These methods are described below.

\subsubsection{Random Forest}
\label{subsec:Random Forest}

Random forests are a popular technique in classical machine learning, due to their predictive ability at a lower computational burden than neural networks \cite{biau2016random}. At their core, random forests are an ``ensemble" learning technique based on decision trees. Ensemble learning is the strategy of averaging predictions from multiple individual models or \textit{estimators}, leading to more robust and accurate predictions. The random forests can be configured to train a predefined number of decision tree estimators for the same training data. Each decision tree makes a target prediction based on training data. Each tree also has a user-specified depth parameter. The depth parameter denotes the number of branches the tree is allowed to create, when fitting to the training data. Typically, increasing depth can increase the predictive capability of the decision tree, as it can learn more intricate features in the data. However, increasing depth beyond a certain limit can also cause over-fitting and reduce accuracy. The precise limit is dependent on the data, and is discovered by trial-and-error. The random forest aggregates the model from all these individual trees, to create an ensemble model.\par

%In this work we select $100$ trees with a maximum tree depth of $50$ for random forests regression, using the Scikit-learn package. This means $100$ decision trees are trained on the same data, but each of them will have slightly different outcomes. The variations are due to the randomness in how each decision tree samples the data, thereby also leading to variations in predictions. After training, the random forests aggregates these variations and makes an ensemble prediction.

\subsubsection{Gradient Boosted Trees}
\label{subsec:Gradient Boosted Trees }

Gradient boosting is another family of ensemble methods fitting a sequence of weak learners (estimator that gives a prediction slightly better than a random guess) on modified versions of the dataset~\cite{natekin2013gradient}. In the Gradient Boosted Tree algorithm, the convergence of the boosting algorithm is improved by computing the gradient of a differentiable loss functions. In Gradient Boosting the base estimator is the Decision Tree estimator and the hyper-parameters in the tuning phase are the number of estimator and the learning rate.

\subsubsection{K-Nearest Neighbors Regression}
\label{subsec:KNN}
The central idea behind the K-nearest neighbors (KNN) is based on the nearest neighbors to query a data point, where $k$ is an integer algorithm parameter. Therefore, the value of a quantity at a point is a weighted average of the $k$ points closest to it~\cite{fix1989discriminatory, altman1992introduction}.  The user specifies the distance metric for computing the weights. There are multiple choices: Uniform, Euclidean, Manhattan, Minkowski etc. With Uniform weights, each neighbor is provided the same weight irrespective of its distance from the query point. In the other distance metrics, the neighbors closer to the query point in that particular space are assigned higher weights than the those further away. Therefore, this metric acts as a weighted average.\par %In this work, we set $k=150$ and choose uniform weights with the Scikit Learn library.\par

\subsubsection{Gaussian Process Regression}
\label{subsec:Gaussian Process}
Gaussian process regression is a non-parametric Bayesian approach towards regression problems. It can capture a wide variety of relations between inputs and outputs by utilizing a theoretically infinite number of parameters and letting the data determine the level of complexity through the means of Bayesian inference \cite{gershman2012tutorial, williams1998bayesian, williams1995gaussian}.

\subsubsection{Multi-layer Perceptron (MLP) Regression}
\label{subsec:MLP}

MLPs are a type of neural network consisting of multiple layers: an input layer, one or more hidden layers, and an output layer. Each layer is fully connected to the next one via non-linear activation functions. Training a neural network on a simulation such that it can be generalized to apply to an experimental dataset that differs from the simulation model in many ways is often challenging. MLPs are particularly susceptible to over-fitting, although there are regularization methods available to counter the problem of over-fitting. Tuning of hyper parameters, such as the activation function, number of hidden layers, number of nodes in each hidden layers, amount of regularization, dropout, enabling/disabling early stopping, and choosing learning rates and optimization strategies are necessary to achieve the best possible performance.

\subsubsection{Convolutional Neural Networks (CNN)}
\label{subsec: Convolutional Neural Networks}
CNNs are a form of neural network in which the linear layers take the form of a set of convolutions~\cite{fukushima1980neocognitron, lecun2015deep, schmidhuber2015deep}. This greatly reduces the number of trainable weights, thereby decreasing the risk of over-fitting, and also allows for computationally efficient implementation. These methods are typically only suitable, however, when the input data has the shift-invariance properties implied by the use of convolutions.

\section{Training Data Generation}
\label{sec:datagen}

The training data for the ML algorithms was generated utilizing GADRAS~\cite{osti_1177049}, incorporating a 145\% relative efficiency HPGe detector with a bismuth side shield and tin filter  using either Pu or U sources in either metal or oxide forms. The sources were contained in one of three geometries, i.e. shells, cylinders, or spheres. Since the ``self-shielding"  is dependent on the source geometry, the $\mathrm{\gammaup}$ spectra are not identical for two identical sources that differ only in geometry. Therefore, an ensemble of training data for each of the respective geometries was generated using a variety of \Uttf\ enrichment/\Puttn isotopic fractions, source thickness, and shielding materials with accompanying thicknesses.  Characteristics of the training data are summarized in Tables~\ref{tab:SummaryGADRASsimU} and~\ref{tab:SummaryGADRASsimPu}.\par
%%%%%%%%%%%%%%%%%%
\begin{table}
\begin{tabular}{|c|c|c|c|c|}
    \hline
    Geometry & No. of & Enrichment & Shielding\\
    & Decks & fraction (\Uttf\ ) & present  \\
    \hline
    Shell & 1800 & 0.000 -- 0.989 & No\\
    \hline   
    Shell & 15839 & 0.003 -- 1.000 & Yes\\
    \hline
    Sphere & 1800 & 0.000 -- 0.995 & No\\
    \hline
    Cylinder & 7000 & 0.000 -- 1.000 & No\\
    \hline
    Cylinder & 20000 & 0.000 -- 1.000 & Yes \\
    \hline
\end{tabular}
\caption{Summary of training data simulations used for uranium in various configurations.}
\label{tab:SummaryGADRASsimU}
\end{table}
\begin{table}
\begin{tabular}{|c|c|c|c|}
    \hline
    Geometry & No. of & Isotopics  & Shielding\\
    & Decks & fraction (\Puttn\ ) & present  \\
    \hline
    Shell & 1800 & 0.000 -- 0.995 & No\\
    \hline   
    Shell & 7920 & 0.230 -- 1.000 & Yes\\
    \hline
    Sphere & 1800 & 0.000 -- 0.995 & No\\
    \hline
    Cylinder & 5000 & 0.560 -- 1.000 & No\\
    \hline
    Cylinder & 20000 & 0.560 -- 1.000 & Yes \\
    \hline
\end{tabular}
\caption{Summary of training data simulations used for plutonium in various configurations.}
\label{tab:SummaryGADRASsimPu}
\end{table}
%%%%%%%%%%%%%%%%%%
For the dataset created with a shell configuration, the thickness of the source shells was between 0.02 and 4 cm. The interior of the shell had a void of radius 1.6 cm for Pu sources with source thickness greater than 2 cm, and for all other cases the outer surface of the source was 6 cm. Spectra generated for sources with spherical geometry had radii ranging from 0.02 to 4 cm. Cylindrical plutonium sources were generated with heights ranging from 0.35 to 1.57 cm and 0.142 to 0.59 cm with corresponding radii ranging from 0.4 to 1.1 cm and 0.353 to 0.931 cm in bare and shielded configurations, respectively. Cylindrical uranium sources were generated with heights ranging from 3.9 to 7.7 cm and 2.45 to 9.76 cm with corresponding radii ranging from 5.35 to 5.44 cm and with 5 to 6 cm radii in bare and shielded configurations, respectively.\par 

For the Shell and Sphere configuration simulations with the shielding material present, iron (Fe), Tantalum (Ta), Polypropylene, or some combination of the aforementioned materials was utilized. The thickness of shielding materials ranged from 1--10, 1--6, and 1--64 cm respectively for the aforementioned materials. The Cylindrical geometry dataset utilized various combination of aluminum (Al), Tantalum (Ta), Iron (Fe), Lead (Pb), and Polypropylene for shielding, while the shielding thickness ranged from 0.05 -- 2 cm. 

\subsection{Comparison of experimental data with GADRAS simulations}
\label{sec:data-sim}
\begin{figure}
    \centering
    \includegraphics[width=0.46\textwidth]{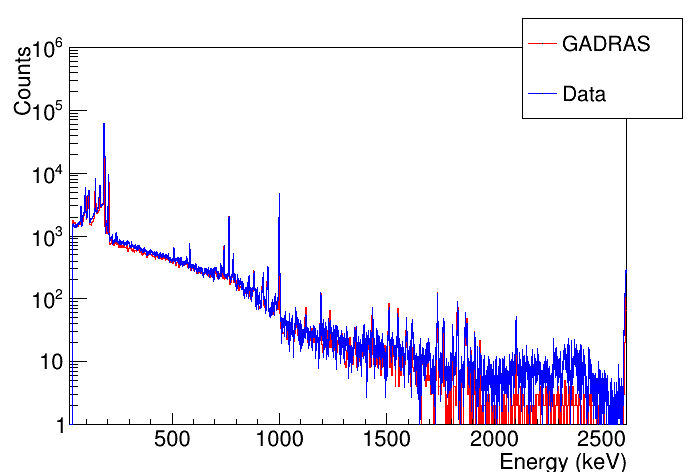}
    \caption{Comparison of 8x binned simulated spectra and background subtracted spectra for a U source, UISO17.
    %\todo{these figures need to be re-done}
    }%
    \label{fig:UISO17}
\end{figure}

\begin{figure}
    \centering
    \includegraphics[width=0.46\textwidth]{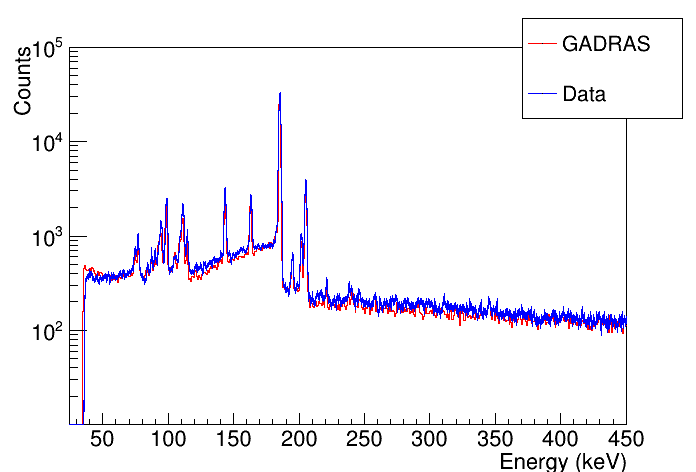}
    \caption{Comparison of 2x binned simulated spectra vs background subtracted spectra for a U source, UISO17.
    %    \todo{redo the figure}
    }%
    \label{fig:UISO17a}
\end{figure}

\begin{figure}
    \centering
    \includegraphics[width=0.46\textwidth]{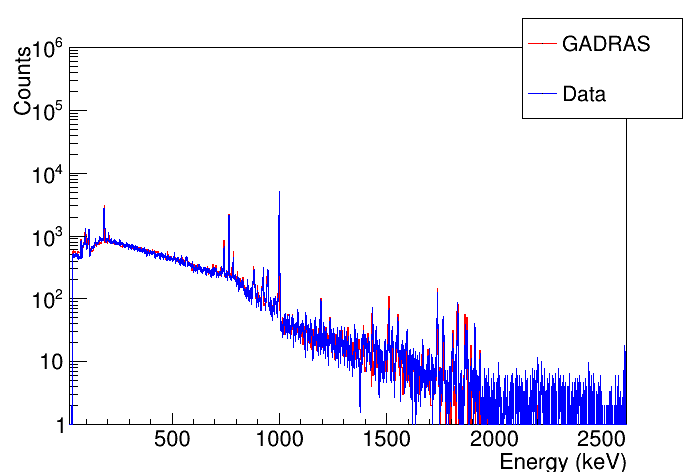}
    \caption{Comparison of 8x binned simulated spectra and background subtracted spectra for a U source, A1127.
    %\todo{these figures need to be re-done}
    }%
    \label{fig:UISO17b}
\end{figure}

\begin{figure}
    \centering
    \includegraphics[width=0.46\textwidth]{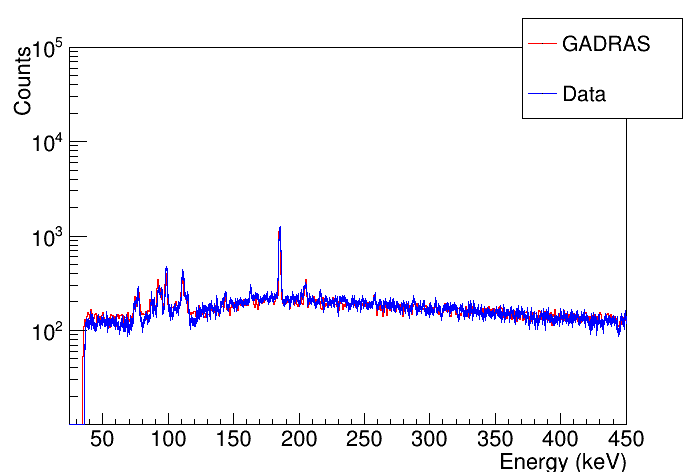}
    \caption{Comparison of 2x binned simulated spectra vs background subtracted spectra for a U source, A1127.
    %    \todo{redo the figure}
    }%
    \label{fig:UISO17c}
\end{figure}

\begin{figure}
    \centering
    \includegraphics[width=0.46\textwidth]{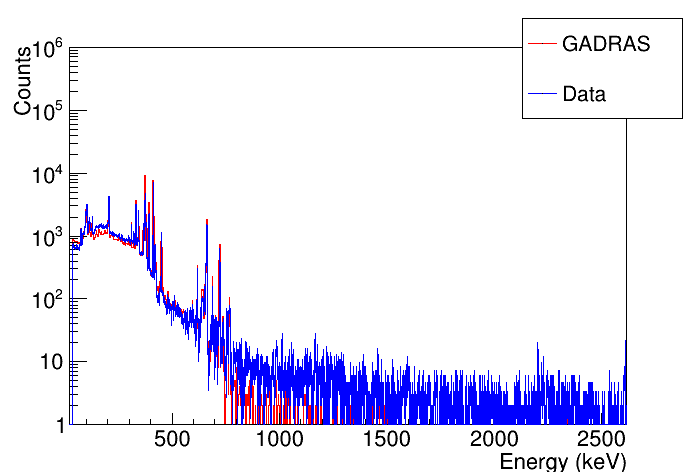}
    \caption{Comparison of 8x binned simulated spectra vs background subtracted spectra for a Pu source, CBNMPu84.}
    \label{fig:CBNMPu84}
\end{figure}

\begin{figure}
    \centering
    \includegraphics[width=0.46\textwidth]{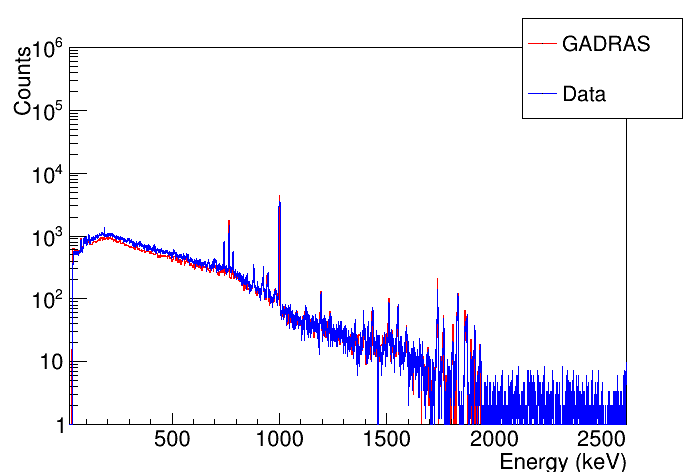}
    \caption{Comparison of 8x binned simulated spectra vs background subtracted spectra for a Du-Shell.}
    \label{fig:DuShell_2xDB}
\end{figure}

In machine learning, the ability of the simulations to replicate the experimental data is
a fundamental issue that must be addressed when simulations are utilized for training and the testing is performed using experimental measurements. To this end, we performed experiments with a HPGe detector using both uranium as well as plutonium sources with and without accompanying shielding and compared these with GADRAS simulated spectra. The simulations were generated using source data sheets for the primary isotopics, geometry, reported age of the material, and dimensions/configurations to model the experimental data. Furthermore, for the GADRAS simulation of \UtOe and \PuOt sources, the mass fractions of uranium and plutonium were adjusted to account for the oxide forms utilized in the experimental data. The isotopic fractions of U/Pu isotopes other than \Uttf, \Utte, \Puttn, and \Putf were chosen based on the certification sheets for the sources. For the depleted uranium shell simulations, generic values of enrichment and miscellaneous isotopic fractions were utilized, while the void/shell thickness were matched to those in the experimental setup. Additionally, for the \UtOe  simulations, $\mathrm{^{40}K}$ and $\mathrm{^{232}Th}$ contents were adjusted to match the background data. The $\mathrm{^{232}U}$ content in the simulation was also adjusted based on the height of the 2614.5 keV photopeak. Some fine tuning in the normalization was performed to match the container material and thickness, where appropriate. The simulation models were run with Poission statistics, and compared to data with terrestrial background contribution subtracted from the spectra. Comparisons of the simulated spectra with the experimental spectra are presented in Figures~\ref{fig:UISO17}--\ref{fig:DuShell_2xDB}.
\section{Pre-processing and feature extraction}
\label{sec:pre}
The experimental analog data pulses obtained from the HPGe detector system, after being converted into digital pulses, are recorded in the units of count per discrete channels. $\mathrm{\gammaup}$ spectra obtained from the GADRAS simulation are also obtained in the units of counts per channel. In either case the counts may be, optionally, pre-processed to remove the continuum background. Detailed discussion on continuum subtraction is provided in subsection~\ref{sec:contsubt}. Additionally, for the experimental data, where terrestrial background is  present, contributions from such background sources are estimated and subtracted from the foreground counts. The net-counts, after optional continuum subtraction and terrestrial background subtraction, are then integrated in a region of interest around photo-peaks of interest to estimate counts associated with each of the photo-peaks. The regions of interests are chosen based on the expected photo-peaks for the two isotopes of U and Pu examined in this study. The mean value of the energy associated with these photo-peaks and their associated net-counts constitute the features for ML training. The impact of the number of features, and the means of reducing the dimensionality of the features during the supervised ML training is investigated in Section~\ref{sec:Dimreduct}. For training samples, the features are accompanied by answer ``keys", which are the relative fraction of \Uttf(\Puttn) with respect to the total fraction of \Uttf and \Utte (\Puttn and \Putf). Here onward, for simplicity, these quantities will be together referred as isotopic ratios or as \Uttf frac and \Puttn frac individually.
\subsection{Continuum Subtraction}
\label{sec:contsubt}

\begin{figure}
    \centering
    \includegraphics[width=0.46\textwidth]{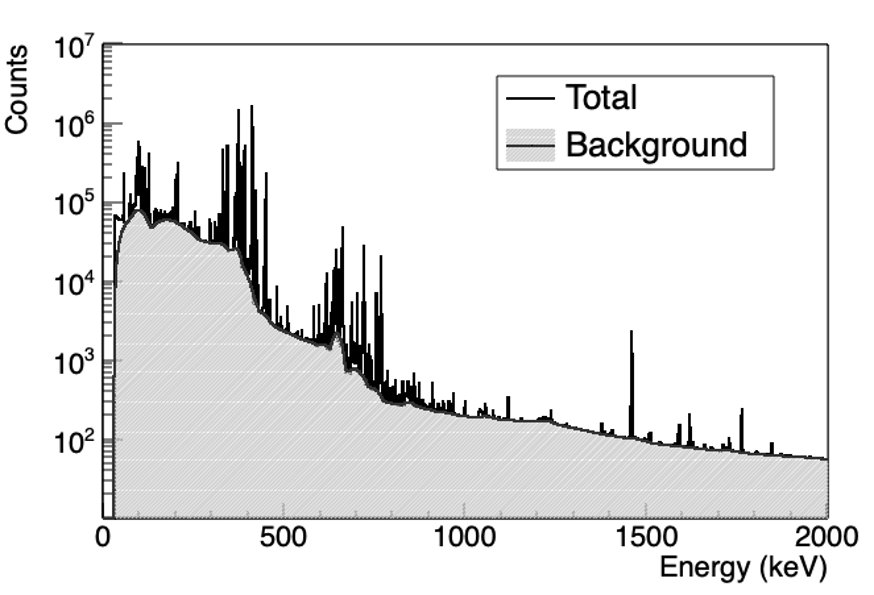}
    \caption{An example of a continuum estimate performed on a $\mathrm{\gammaup}$ spectra for a Pu source.}
    \label{fig:ContinuumSubt}
\end{figure}

Subtraction of the continuum background produced from scattering were examined to understand the impact on the isotopic determination. As such, a Sensitive Nonlinear Iterative Peak (SNIP) clipping algorithm implemented in TSpectrum class of ROOT framework~\cite{MORHAC1997113, MORHAC1997385, 2000NIMPA.443..108M} was utilized for one-dimensional background estimation. The number of iterations was examined in estimating the continuum. The optimal number of iterations was chosen to be 20 based on the ability to remove adequate amount of continuum without resulting in negative counts in the subtracted spectra.  An example spectra with the continuum background estimate with this method is provided in Figure~\ref{fig:ContinuumSubt}.

\section{Investigations using simulated data}
\label{sec:sim}
The determination of the isotopic content of uranium or plutonium is a complex function of numerous factors including: the source geometry, source thickness, shielding material composition, shielding thickness, possible inherent impurities e.g. $\mathrm{^{232}U}$ along with the isotopic ratios of the isotopes in question i.e. \Uttf\ or \Puttn\ . Furthermore, the accuracy in determining the isotopic fraction is determined by the ability to adequately sample these variables in the training set as well as the representativeness of the training data to the testing data, the quantity of training data from which to learn, and the ability to adequately train the given ML algorithm.

%\section{Hyper-parameter tuning}
%\label{sec:Hyper}
%In applying the various ML algorithms to determine the isotopics of either Pu or U there are numerous parameters that can be tuned. Table xx summarizes the hyper-parameters examined for each of the respective ML algorithms.

%The same test was performed for MLP algorithm tuning the various hyper-parameters available in Scikit Learn package. Scanning the results for wide range of hyper-parameters (type of activation function used, strength of regularization, number of hidden layers, number of nodes in each hidden layers, solver, learning rate), the minimum value of mean error on the isotopics determination was roughly 0.06 with this algorithm. As mentioned in Section~\ref{subsec:MLP}, neural networks are particularly susceptible to over-training. 
Since the final goal of this study is to apply the ML algorithm to experimental data that may differ significantly from the training sample in multiple different ways, i.e. amount of shielding present, source geometry, background spectra, etc., the algorithm needs to be robust against over-fitting. \par

\subsection{Bare: No Shielding}
As an initial test, two simulated datasets (Spheres and Cylinders) were utilized to examined the ability of different ML algorithms to predict the isotopic ratios for both Pu and U with no shielding materials present. This test represents the most simplistic mapping from the spectra to isotopic ratios that can be learnt.  That is, no alteration of the line intensities due to the shielding needs to be learned. A sample result showing the absolute mean error ($\mathrm{\abs{true - predicted}}$) and the standard deviation of the error for training and testing with a dataset generated using cylindrical geometry is presented in Table~\ref{tab:finalbarecylinder_before}.

\begin{table}
\resizebox{\columnwidth}{!}{%
\begin{tabular}{|c|c|c|}
    \hline
    %\textbf{Method} & \textbf{\Uttf/\Utte} & \textbf{\Puttn/\Putf}\\
    \textbf{Method} & \textbf{\Uttf} frac & \textbf{\Puttn} frac\\
    \hline
    Nearest & 0.0012	$\pm$	0.0041& 0.0052	$\pm$	0.0044\\
    \hline
    Decision & 0.0014	$\pm$	0.0016& 0.0021	$\pm$	0.0027\\
    \hline
    Random & 0.0009	$\pm$	0.0009&  0.0012	$\pm$	0.0016\\
    \hline
    GB & 0.0023	$\pm$	0.0018& 0.0027	$\pm$	0.0025\\
    \hline
    Gaussian* & 0.0007	$\pm$	0.0007& 0.0009	$\pm$	0.0007\\
    \hline
    FCNN* & 0.0007	$\pm$	0.0006& 0.0006	$\pm$	0.0006\\
    \hline
\end{tabular}
}
\caption{Mean error and the standard deviation of error in the isotopic ratios using cylinder simulations with no shielding materials. Results marked as `*' were produced using algorithm implemented in Mathematica.}
\label{tab:finalbarecylinder_before}
\end{table}
\par
Examination of results of the bare geometries, example shown in Table~\ref{tab:finalbarecylinder_before}, indicates that all of the ML methods with the chosen parameter settings perform an excellent job at predicting the isotopics. This is to be expected since the ratio of the line intensities is solely a function of the thickness of the radioisotopes. One additional finding from the analysis of these datasets is that the ML algorithms are able to adequately treat the continuum and therefore remove the time consuming continuum subtraction step.  However, to quantify the ability of the ML algorithms to perform this function we utilized a continuum subtraction algorithm, as outlined in Section~\ref{sec:contsubt}. An example result is provided in Table~\ref{tab:finalbarecylinder_after}. More information on the hyperparameter examination is provided in Section~\ref{sec:Hyper}.\par
\begin{table}
\resizebox{\columnwidth}{!}{%
\begin{tabular}{|c|c|c|}
    \hline
    \textbf{Method} & \textbf{\Uttf} frac & \textbf{\Puttn} frac\\
    \hline
    Nearest & 0.0013	$\pm$	0.0042 & 0.0051	$\pm$	0.0045\\
    \hline
    Decision & 0.0020	$\pm$	0.0028 & 0.0031	$\pm$	0.0030\\
    \hline
    Random & 0.0011	$\pm$	0.0012 & 0.0017	$\pm$	0.0016\\
    \hline
    GB & 0.0024	$\pm$	0.0022 & 0.0033	$\pm$	0.0036\\
    \hline
    Gaussian* & 0.0010 $\pm$ 0.0016 & 0.0020 $\pm$ 0.0019\\
    \hline
    FCNN* & 0.0011 $\pm$ 0.0009 & 0.0012 $\pm$ 0.0013\\
    \hline
\end{tabular}
}
\caption{Mean error and the standard deviation of error in the isotopic ratios using simulations with no shielding materials after the simulation was pre-processed to subtract the continuum. Results marked as `*' were produced using algorithm implemented in Mathematica.}
\label{tab:finalbarecylinder_after}
\end{table}
Examination of Table~\ref{tab:finalbarecylinder_before} and \ref{tab:finalbarecylinder_after} indicates that the ML algorithms indeed perform well in removing the continuum. The slight decrease in performance upon separate continuum subtraction may be attributed to the decrease in statistics, and the uncertainty in continuum subtraction procedure. 

\subsection{Testing the impact of shielding}
The previous investigations did not include any shielding.  It is instructive to examine the ability of the learning algorithms to learn a much more complex multi-dimensional function i.e. determine the isotopic ratios of \Uttf\ and \Puttn\ when different shielding materials with different thicknesses are present.  Indeed, as may be observed from examination of Table~\ref{tab:finalshieldedcylinder_before}, \ref{tab:finalshieldedcylinder_after} errors increase relative to those obtained without shielding.\par
\begin{table}
\resizebox{\columnwidth}{!}{%
\begin{tabular}{|c|c|c|}
    \hline
    \textbf{Method} & \textbf{\Uttf} frac & \textbf{\Puttn} frac\\
    \hline
    Nearest & 0.0636	$\pm$	0.0684  & 0.0253	$\pm$	0.0256\\
    \hline
    Decision &  0.0370	$\pm$	0.0492 & 0.0060	$\pm$	0.0078\\
    \hline
    Random & 0.0248	$\pm$	0.0322 & 0.0032	$\pm$	0.0040\\
    \hline
    GB & 0.0264	$\pm$	0.0317 & 0.0054	$\pm$	0.0056\\
    \hline
    Gaussian* & 0.0290	$\pm$ 0.0290 & 0.0012	$\pm$ 0.0012\\
    \hline
    FCNN* & 0.0310	$\pm$	0.0290 & 0.0083 	$\pm$	0.0084\\
    \hline
\end{tabular}
}
\caption{Mean error and the standard deviation of error in the isotopic ratios using simulations with shielding materials. Results marked as `*' were produced using algorithm implemented in Mathematica.}
\label{tab:finalshieldedcylinder_before}
\end{table}

\begin{table}
\resizebox{\columnwidth}{!}{%
\begin{tabular}{|c|c|c|}
    \hline
    \textbf{Method} & \textbf{\Uttf} frac & \textbf{\Puttn} frac\\
    \hline
    Nearest & 0.0503	$\pm$	0.0539 & 0.0237	$\pm$	0.0255\\
    \hline
    Decision &  0.0200	$\pm$	0.0222 & 0.0075	$\pm$	0.0082\\
    \hline
    Random & 0.0134	$\pm$	0.0161 & 0.0039	$\pm$	0.0043\\
    \hline
    GB & 0.0148	$\pm$ 0.0141 & 0.0064	$\pm$	0.0052\\
    \hline
    Gaussian* & 0.0148 $\pm$ 0.0145 & 0.0026	$\pm$	0.0023\\
    \hline
    FCNN* & 0.0120	$\pm$ 0.0110 & 0.0035	$\pm$	0.0032\\
    \hline
\end{tabular}
}
\caption{Mean error and the standard deviation of error in the isotopic ratios using simulations with shielding materials after the simulation was pre-processed to remove continuum background. Results marked as `*' were produced using algorithm implemented in Mathematica.}
\label{tab:finalshieldedcylinder_after}
\end{table}

Examination of Table~\ref{tab:finalshieldedcylinder_before} and \ref{tab:finalshieldedcylinder_after} also reveals that the impact of background subtraction has a minimal impact on the errors.  All ML methods, with the possible exception of the nearest neighbor, appear to offer comparable performance. Finally, we observe that in the dataset with shielding applied, the plutonium predictions are significantly better than those for uranium. Shielding adds extra scattering background to the observed spectra, which makes the ratio of the photo-peak counts to the scatter background smaller. Most of the photo-peaks features that are useful for the uranium enrichment determination are far apart in energy, with different amount of scatter present under the peaks. Furthermore, these photo-peaks are often also in the low energy region, where photo-peak to continuum background ratio is already smaller than for photo-peaks in medium energy range, which are more useful for plutonium isotopic determination. Therefore, accuracy in the continuum background determination, whether it is through a separate step applied during pre-processing or one done automatically by the ML algorithm, impacts the uranium enrichment estimate asymmetrically as compared to the plutonium isotopics determination.\par

\subsection{Generalization of ML Algorithms}
In nuclear safeguards applications, many field parameters, such as source geometry and shielding material properties, are unknown. The previous investigations reported results for cases in which the training and testing datasets were drawn from the same general population e.g. training and testing on cylinders or other common geometries; or with common shielding materials and thickness and fixed geometries. A common issue in ML is the ability of a given ML algorithm, with a given training set, to generalize e.g. to make predictions using testing data that may be different in either a known or unknown manner from the training data. Testing the validity of the ML algorithm's performance with data that differs from the training sample in either source geometry or shielding materials allows for identification and quantification of possible sources of uncertainty.  In the first investigations, training with one geometry and testing on another was examined. It was observed that the training with bare spheres and testing on bare cylinders resulted in significantly worse performance than those results obtained above, with  mean errors on the order of 0.10--0.15 for most algorithms. An additional investigation in which training with shielded shells and testing with shielded cylinders revealed even higher degradation in the performance for all of the ML algorithms owing to the increase in complexity and difference between the phase space of the training and testing samples. To illustrate the second issue, ML algorithms were initially trained on simulations with cylindrical geometry generated without shielding materials and tested on simulations with shielding materials. The predictive ability and generalization ability of the ML algorithms was degraded as reflected by mean error values in the range of 0.05--0.10 for Pu and 0.10--0.15 for U.  The process was later repeated with the training and testing populations swapped. The mean absolute errors obtained were $<$ 0.01 for most ML algorithms. The lower value of mean absolute error when training on a sample that was produced with shielding materials ranging in material type and thickness implies that increasing the heterogeneity in the training sample to widen the physics phase space increases the overall generalization ability of the algorithm, as predicted. 

To address the degradation in the performance when a variety of geometries may be present all of the training data was combined.  The results of these investigations are provided in Tables~\ref{tab:allsim_before} and \ref{tab:allsim_after}  with and without continuum subtraction, respectively.

\begin{table}
\resizebox{\columnwidth}{!}{%
\begin{tabular}{|c|c|c|}
    \hline
    \textbf{Method} & \textbf{\Uttf} frac & \textbf{\Puttn} frac\\
    \hline
    Nearest & 0.0120	$\pm$	0.0310  & 0.0190	$\pm$	0.0270\\
    \hline
    Decision &  0.0073	$\pm$	0.0140 & 0.0087	$\pm$	0.0240\\
    \hline
    Random & 0.0038	$\pm$	0.0080 & 0.0056	$\pm$	0.0160\\
    \hline
    GB & 0.0170	$\pm$	0.0160 & 0.0150	$\pm$	0.0200\\
    \hline
    Gaussian* & 0.1300 $\pm$ 0.1100 & 0.0093 $\pm$	0.0095 \\
    \hline
    FCNN* & 0.1500 $\pm$ 0.1300 & 0.0350  $\pm$	0.0250 \\
    \hline
    CNN* & 0.2300 $\pm$ 0.1600 & 0.0300  $\pm$	0.0230 \\
    \hline
\end{tabular}
}
\caption{Mean error and the standard deviation of error in the isotopic ratios using all simulations. Results marked as `*' were produced using algorithm implemented in Mathematica.}
\label{tab:allsim_before}
\end{table}

\begin{table}
\resizebox{\columnwidth}{!}{%
\begin{tabular}{|c|c|c|}
    \hline
    \textbf{Method} & \textbf{\Uttf} frac & \textbf{\Puttn} frac\\
    \hline
    Nearest & 0.0130	$\pm$	0.0330  & 0.0190	$\pm$	0.0290\\
    \hline
    Decision &  0.0076	$\pm$	0.0190 & 0.0085	$\pm$	0.0210\\
    \hline
    Random & 0.0041	$\pm$	0.0081 & 0.0048	$\pm$	0.0150\\
    \hline
    GB & 0.0180	$\pm$	0.0170 & 0.0130	$\pm$	0.0180\\
    \hline
    Gaussian* & 0.0050  $\pm$ 0.0060 & 0.0050 $\pm$	0.0090 \\
    \hline
    FCNN* & 0.0270  $\pm$ 0.0210 & 0.0090  $\pm$	0.0070 \\
    \hline
    CNN* & 0.0540 $\pm$ 0.0460 & 0.0140  $\pm$	0.0110 \\
    \hline
\end{tabular}
}
\caption{Mean error and the standard deviation of error in the isotopic ratios using all simulations after subtracting contributions from continuum.  Results marked as `*' were produced using algorithm implemented in Mathematica.}
\label{tab:allsim_after}
\end{table}

Examination of Table~\ref{tab:allsim_before} and Table~\ref{tab:allsim_after} reveals excellent performance of the ML algorithms, without continuum subtraction, in determining the Pu isotopic content.  However, for the uranium isotopic content, it was found that the Gaussian processes and the neural networks did not perform adequately.  Examinations of the isotopic content predictions using all ML algorithms revealed excellent performance when a separate continuum subtraction was done during pre-processing.

Finally, the isotopics for all of the simulated data for plutonium and uranium were evaluated using FRAM.  The results were significantly worse 0.074 for plutonium and 0.11 for uranium than those obtained via the machine learning algorithms.

\section{Hyper-parameter Examinations}
\label{sec:Hyper}
\begin{table}
\centering
\resizebox{\columnwidth}{!}{%
\begin{tabular}{|l|l |c |r|}
\hline
Method    & Parameters & Range/             & Value/       \\
          &            & Methods            & Range        \\
          &            & Explored           & Selected      \\\hline
Nearest   & Neighbors: & 1--32000           & 1             \\
Neighbors & Methods:   & KDtree/             & Auto               \\
          &            & Brute/Auto          &               \\
          & Distance:  & Uniform/            & Minkowski     \\ 
          &            & Chebyshev/          & (Euclidean)   \\
          &            & Euclidean/          &                \\
          &            & Manhattan/          &                \\
          &            & Minkowski          &                \\\hline
Decision  & Max Depth: & 10--100            & 50             \\
Tree      & Splitter:  & Best               & Best           \\
          & Loss:      & MSE                & MSE             \\
          & Feature       &                    &                 \\ 
          & Fraction:     &  0.05--1           & 1               \\\hline
Random    & No. of trees: & 10--100            & 100             \\
Forest    & Leaf Size:    & Unlimited          & Unlimited      \\
          & Max Depth:    & 10--100/None       & 50             \\
          & Loss:         & MSE                & MSE             \\
          & Feature       &                    &                 \\ 
          & Fraction:     & 0.05--1           & 1               \\\hline
Gradient  & No. of trees: & 10--300            & 200             \\
Boosted   & Leaf Number:  & 5--50/None         & None      \\
Trees     & Max Depth:    & 2--25              & 4             \\
          & Min samples   &                    &              \\
          & for split:    & 2--10              & 5                 \\ 
          & Loss:         & MSE                & MSE             \\
          & Learning      &                    &                 \\ 
          & Rate:         & 0.01--0.4         & 0.1               \\
          & Feature       &                    &                 \\ 
          & Fraction:     & 0.1--1            & 1               \\\hline
Fully     & Layers:       & 2--10              & 2              \\
Connected & Activation:   & SELU/Tanh          & Tanh      \\
Neural    & No of params: & 15250--100000      & 10250             \\
Network   & DropOut:      & 0--0.1              &0.01              \\
          & Epochs:       & 100--1000          & 100                 \\ 
          & Optimization  &                    & MSE             \\
          & Method:       & ADAM/SGD           & ADAM               \\
          & Learning      &                    &                 \\ 
          & Rate:         & 0.001--0.1        & 0.01               \\\hline
Convol-   & Layers:       & 2--10              & 2              \\
utional   & No of params: & 10250--100000      & 35324             \\
Neural    & DropOut:      &0--0.1              & 0.1937              \\
Network   & Activation:   & SELU/ Tanh          & SELU      \\
          & Epochs:       & 100--1000          & 100                 \\ 
          & Optimization  &                    &                 \\
          & Method:       & ADAM/SGD/          & Logistic\\
          &               & LogisticSigmoid    & Sigmoid\\
          & L2:  & 0--0.1             & 0.01 \\
          & Learning      &                    &                 \\ 
          & Rate:         &  0.001--0.1        & 0.001               \\   \hline   
\end{tabular}%
}
\caption{Hyperparameters tested and selected for different ML algorithms.}
\label{tab:hyperparam}
\end{table}
The parameters that need to be defined prior to training a ML algorithm are commonly termed as hyperparameters. There is currently no known method to determine which hyperparameters have an impact on model performance before training.  Consequently, for each of the respective methods  a range of hyperparameters was explored.  Furthermore, because the objective of this work is to train models using simulation data and test using experimental data, pre-cautions were taken to avoid over-fitting.  A summary of the hyper-parameters examined, and the parameters utilized for subsequent investigations, for each of the respective methods is provided in Table~\ref{tab:hyperparam}.

\section{Experimental Data and Results}
\label{sec:Exp}

The previous analyses were performed using simulated spectra generated using GADRAS for both training as well as testing.  In this Section, we explore the use of the simulations for training and experimental data for testing.  The details of the experimental configurations are outlined in  Section ~\ref{sec:ExpDesc} and the application of ML algorithms are presented in Section~\ref{sec:Results}.  However, before presenting these details, we note that at the time of experimental data collection, terrestrial background data is taken with identical settings to the experimental data. Often the background files are generated with longer collection times than the experimental data so as to minimize the effects of statistical fluctuation when subtracting the terrestrial background counts from the foreground counts. Prior to subtraction from the foreground counts, the background counts are scaled accordingly based on the relative count time for the background file with respect to the count time for the experimental data.\par
We note that in the GADRAS simulations, the spectra do not include contributions from terrestrial background; hence, this process is not applicable for simulation. 

%However, a set of $\mathrm{\gammaup}$ spectra were simulated to describe possible background contribution in the real data. This type of background is largely variable and unpredictable since it is dependent on many parameters and conditions surrounding the experiment or real-life application scenario. Multiple background simulation files are added to the training dataset, along with signal data, in order to ensure that ML algorithms are able to estimate an absence of signal in an experimental dataset that has no source present. %Therefore, validation studies are performed to ensure that the ML algorithm is able to estimate an absence of signal in the simulated background only spectra. 

\subsection{Experimental Description}
\label{sec:ExpDesc}
Experimental dataset with multiple source and shielding configurations, source geometries, and source forms were utilized to enable the testing of the ML algorithms. Configurations included bare and shielded cans of uranium and plutonium oxide with a wide range of isotopics, depleted uranium spheres and shells (both bare and shielded), and plutonium spheres with various shielding materials and thicknesses. 

\subsubsection{Uranium and plutonium oxide sources}
The uranium (\UtOe) and plutonium (\PuOt) oxide dataset analyzed were collected with an ORTEC Detective X and LANL Detector S respectively. The Detective X is a handheld, mechanically cooled HPGe detector with 50\% relative efficiency. The Detective X has a range of 8 MeV with $2^{14}$ channels.  The LANL Detector S is an ORTEC poptop liquid-nitrogen cooled HPGe detector with a relative efficiency of roughly 140\%. This detector has a range of 12 MeV with $2^{15}$ channels. The Detector S response function was incorporated into the GADRAS simulations as detailed in Section~\ref{sec:datagen}. At the time of data collection, Detector S incorporated a bismuth side shield to reduce the background radiation contribution to the measured spectra. Additionally, a thin front filter made of tin was also present to filter out low energy photons. 

Fourteen data sets, seven without shielding and seven with shielding material present, were collected for both the uranium and plutonium oxide sources. Uranium enrichment and plutonium isotopics were in the range of 0.7--91.3\% and 63.2(25.7) -- 98.0(2.0)\% \Puttn(\Putf)  respectively. Additional reported isotopes are listed in Tables \ref{tab:U3O8Iso} and \ref{tab:MiscPuO2Iso}.  The uranium oxide samples were approximately 1 kg whereas the plutonium samples had mass between 1.6--5.8 g. Thin sheets of lead were used for the shielded measurements, with a thickness of 3.175 mm for the uranium and 4 mm for the plutonium. A complete set of plutonium oxide measurements were collected for 300 seconds at a source-to-detector distance of 50 cm. The uranium oxide measurements was taken with better counting statistics:  600 seconds at a source-to-detector distance of 25 cm, with an exception of an unshielded 91\% enriched oxide measurement, which performed at a source-to-detector distance of 50 cm to ensure an acceptable dead time in the detector. 
\par

\begin{table}
\resizebox{\columnwidth}{!}{%
\begin{tabular}{|c|c|c|c|}
    \hline
    \textbf{\%U234} & \textbf{\%U235} & \textbf{\%U236} & \textbf{\%U238}\\
    \hline
     0.005--0.910 & 0.716--91.340 & 0.002--0.335 & 7.417--99.277\\
    \hline
\end{tabular}
}
\caption{Range in weight \% of reported isotopes relative to total U for \UtOe cylindrical sources dated 9/6/1988.}
\label{tab:U3O8Iso}
\end{table}

\begin{table}
\resizebox{\columnwidth}{!}{%
\begin{tabular}{|c|c|c|c|}
    \hline
    \textbf{\%Pu238} & \textbf{\%Pu241} & \textbf{\%Pu242} & \textbf{\%Am241}\\
    \hline
     0.002--1.177 & 0.014--5.693 & 0.003--4.239 & 0.009--2.510\\
    \hline
\end{tabular}
}
\caption{Range in weight \% of miscellaneous isotopes relative to total Pu for \PuOt cylindrical sources dated 1/1/1990.}
\label{tab:MiscPuO2Iso}
\end{table}

\subsubsection{Depleted uranium shell data}
The depleted uranium (DU) measurements were performed using a LANL Detector S, described above, with a bismuth collimator and tin front filter. Fifty-four data sets were taken, six in nine different configurations at a source to detector distance of 1m. The configurations utilized various combinations of three stacked DU shells of 6.35 mm thickness, while keeping the outer diameter of the DU shells at 15.24 cm. Some configurations were taken without any shielding, and some utilized shielding from either one or two stacked aluminum shells. The aluminum shells were 1.27 cm in thickness.
\subsubsection{BeRP ball data}
The BeRP (Beryllium Reflected Plutonium) ball~\cite{berpballdata} data was collected with Detector K, a 140\% relative efficiency liquid-nitrogen-cooled HPGe detector that is similar to Detector S, at the Nevada Nuclear Security Site. The BeRP ball is a sphere of 7.59 cm diameter alpha-phase plutonium clad with a 0.3 mm of SS304, and weighs 4.48 kg~\cite{SS304}. Although present in the name of the object, the original beryllium reflector was not used in these configurations. The dataset collected was taken at a source-to-detector distance of 50 cm, both unshielded and with shielding (polyethylene)  of thicknesses between 2.54-10.16 cm in conjunction with other combinations of shielding materials such as nickel, steel, mock high explosives, and aluminum ranging in total thickness from 1.27-7.62 cm.
%\subsubsection{Terrestrial background}
%\label{sec:terrbkg}
%File: FRAM U235\% Results of Data and Simulation.
%\begin{tabular}{|c|c|c|c|}
%    \hline
%    \textbf{Source} & \textbf{Data} & \textbf{Simulated} & \textbf{Reported} \\
%    \hline
%    A-323-1 & 13.73 & 6.31 & 10.1 \\
%    \hline
%    A1-1126-2 & 3.51 & 3.86 & 3.1 \\
%    \hline
%    A1-1127-2 & 0.903 & 1.005 & 0.723 \\
%    \hline
%    UISO-17 & 18.9 & 13.0 & 17.5\\
%    \hline
%    UISO-38 & 35.1 & 33.9 & 38 \\
%    \hline
%    UISO-66 & 62.9 & 61.5 & 66.4 \\
%    \hline
%    UISO-91 & 91.9 & 90.2 & 91.4 \\
%    \hline
%\end{tabular}
%\\

\subsection{Dimensionality Reduction}
\label{sec:Dimreduct}

The HPGe detectors utilized in these investigations have 16384 (Detector X) and 32768 (Detector (S) channels. To reduce the dimensionality of the features for which the ML algorithms were trained, we selected a total of 172 features based on the emission lines of the isotopes under investigation.  This dimensionality reduction was performed due to established observation that  when training a ML algorithm in a large multi-dimensional space, there are often redundant features that add noise to the dataset, without improving the performance of the algorithm. Further investigations into improving ML algorithm performance were performed by applying additional dimensionality reduction using two approaches: 1) physics based feature reduction, and 2) Principle component analysis (PCA).\par
For the physics based feature reduction, we select 9 and 10 prominent $\mathrm{\gammaup}$ peaks for U and Pu data/simulations, respectively. The selections were made based on the most commonly used photo-peaks in $\mathrm{\gammaup}$ spectroscopy for plutonium and uranium. Comparison of the results from this method of dimensionality reduction did not improve the mean absolute error and the standard deviation in simulation test dataset. Similarly, the PCA based method also did not reduced the absolute error in a systematic way.\par
As discussed in Section~\ref{sec:data-sim}, since we utilized a simulation model for constructing a training dataset, there is a potential for biased results due to model dependence. This potential bias was examined by inspecting the spectra generated with a GADRAS model for a sampling of the experimental dataset with known parameters. After observing larger disagreements in the lower and higher energy ranges, the number of features was reduced to include features only in the 100--1000 keV range. Although this improved the mean absolute error, this type of ad hoc dimensionality reduction cannot be generalized without the knowledge of the source of data-simulation discrepancy.\par

\subsection{Results} 
\label{sec:Results}
To investigate the performance of ML algorithms using the experimental data discussed in Section~\ref{sec:ExpDesc}, five algorithms: Decision Trees, MLP, Gradient Boosted Trees, Nearest Neighbors and Random Forests, were considered. The results were compared with results obtained using FRAM software. For the uranium dataset, the comparisons were performed using both the `HEU' and `LEU' models.\par
\begin{figure}
   \centering
    \includegraphics[width=0.45\textwidth]{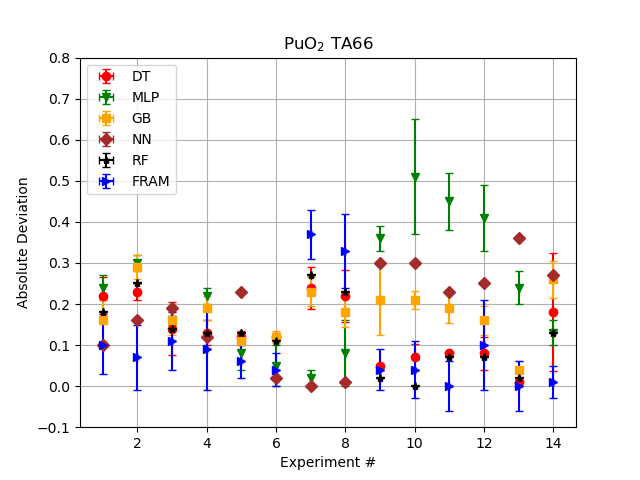}
    \includegraphics[width=0.47\textwidth]{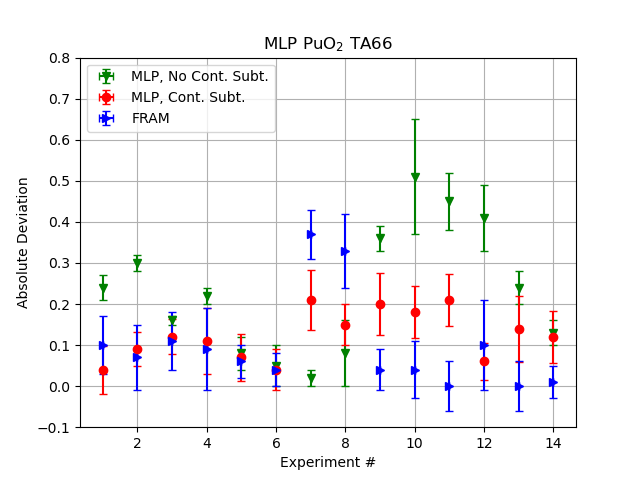}
    \caption{Comparison of absolute deviation from true isotopics ratio for Pu oxide data for various ML algorithms and FRAM (top), and MLP with and without continuum subtraction and FRAM (bottom). Combined statistical and systematic uncertainties are reported at 1 $\sigma$ for all the algorithms except for Random Forest and Nearest Neighbors. Error bars in FRAM results are the 'sigma' values returned by the FRAM software.}
    \label{fig:PuO2}
\end{figure}

The results obtained using the small scale plutonium oxide sources are presented in Figure~\ref{fig:PuO2}.  The error bars for results, provided in Figure~\ref{fig:PuO2}, include combined statistical and systematic uncertainties for all the methods except for Decision Trees and Nearest Neighbors (these methods were found to have very low errors due to the lack of systematic uncertainties which were found to be the dominant source of error). The statistical uncertainties were estimated by varying the photo-peak counts with a poisson model and repeating the ML algorithm implementation for each instance of the variation. The systematic uncertainties account for the variation in the ML results when repeating the training and testing with identical conditions and parameters and incorporating a different random seed for algorithm initialization. Figure~\ref{fig:PuO2} (top) shows that no single ML algorithm outperforms the others for all 14 experiments considered; however, the MLP and Nearest Neighbor methods were found to perform better than the conventional method in a few of the experimental cases. Once the data is pre-processed to remove the continuum, in general the MLP algorithm performs comparable to the conventional method, within the uncertainties of both methods, as shown in Figure~\ref{fig:PuO2} (bottom).\par
\begin{figure}
   \centering
    \includegraphics[width=0.45\textwidth]{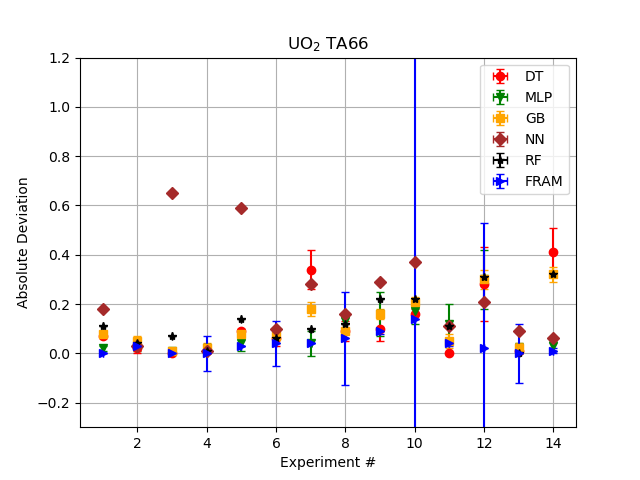}
    \includegraphics[width=0.47\textwidth]{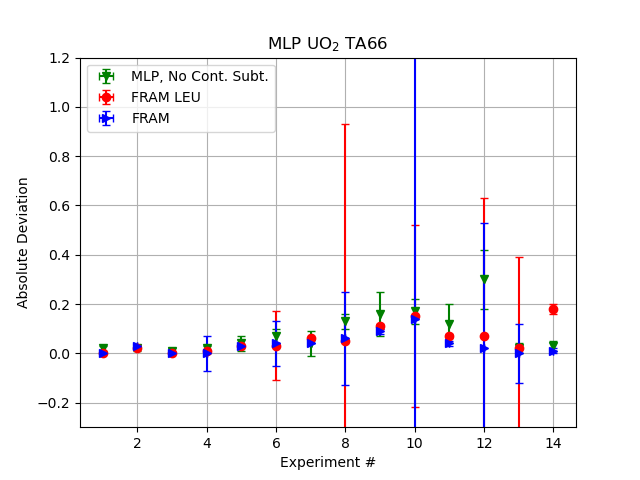}
    \caption{Comparison of absolute deviation from true isotopics ratio for U oxide data for various ML algorithms and FRAM HEU model (top), MLP vs FRAM LEU model (bottom, red) and FRAM HEU model (bottom, blue). Combined statistical and systematic uncertainties are reported at 1 $\sigma$ for all the algorithms except for Random Forest and Nearest Neighbors. Error bars in FRAM results are the 'sigma' values returned by the FRAM software.}
    \label{fig:UO2}
\end{figure}

The uranium oxide results, as provided in Figure~\ref{fig:UO2} (top), show consistently smaller absolute deviations for the MLP as compared to other ML algorithms. The bottom Figure~\ref{fig:UO2} shows that although the results from MLP method are comparable to FRAM results, the estimated uncertainties in some cases (experiment numbers: 6, 8, 10, 12 and 13) are much smaller for the ML method than for the conventional approach.\par
\begin{figure}
   \centering
    \includegraphics[width=0.45\textwidth]{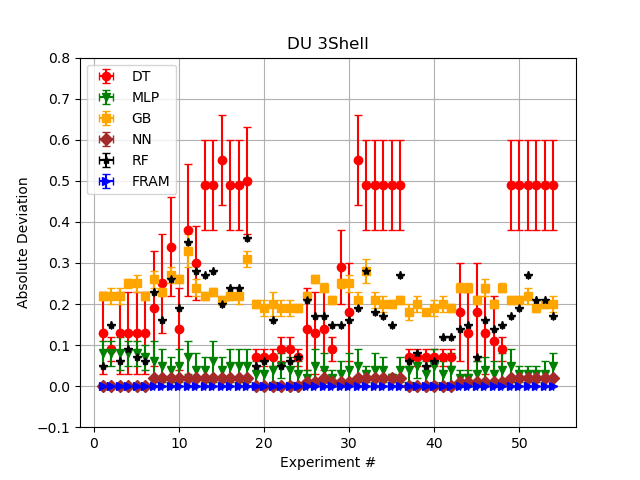}
    \includegraphics[width=0.45\textwidth]{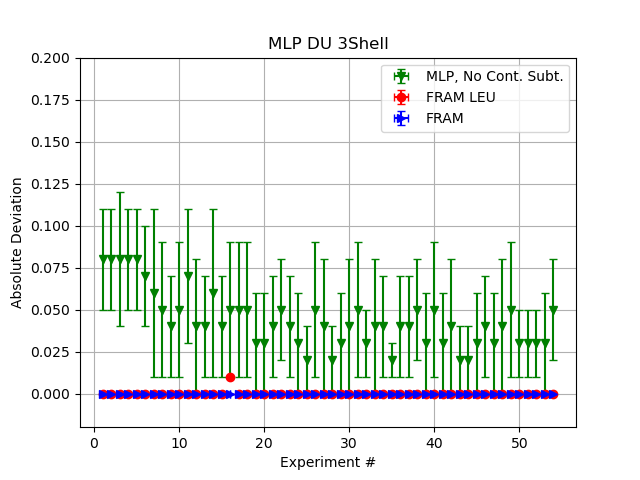}
    \caption{Comparison of absolute deviation from true enrichment value for depleted uranium data for various ML algorithms and FRAM HEU model (top), MLP vs FRAM LEU model (bottom, red) and FRAM HEU model (bottom, blue). Combined statistical and systematic uncertainties are reported at 1 $\sigma$ for MLP, Decision Trees, and Gradient Boosted Trees. Error bars in FRAM results are the 'sigma' values returned by the FRAM software.}
    \label{fig:DuShell}
\end{figure}

The results for depleted uranium shell in Figure~\ref{fig:DuShell} (top) show that decision tree based methods do not perform as well as the Nearest Neighbor and MLP methods. Absolute deviation for the Nearest Neighbor method are comparable to the conventional method for most experiments. The good performance of Nearest Neighbor method is perhaps due to the inclusion of a large number of simulations with enrichment close to that of the depleted uranium in shell configurations. Since this method relies on finding sets of training data points closest in distance to the query, and taking an average of the closest solutions, having a well represented training sample is expected to enhance the performance. The MLP method, despite having larger uncertainty in the estimate as compared to the traditional approach, shows mostly small mean absolute deviation of $<0.05$ for most experiments as shown in Figure~\ref{fig:DuShell} (bottom).\par
\begin{figure}
   \centering
    \includegraphics[width=0.45\textwidth]{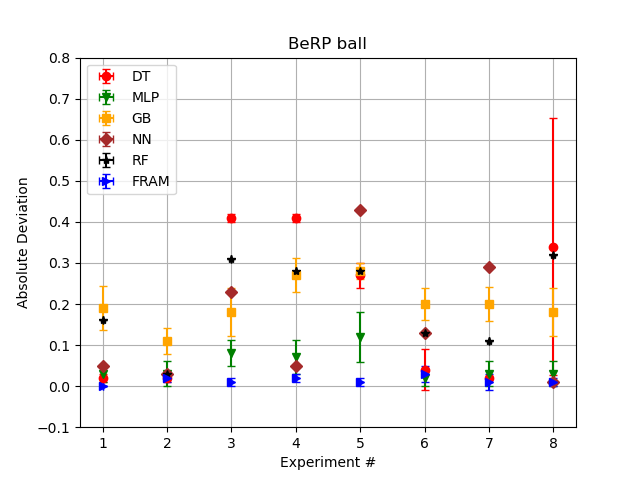}
    \includegraphics[width=0.45\textwidth]{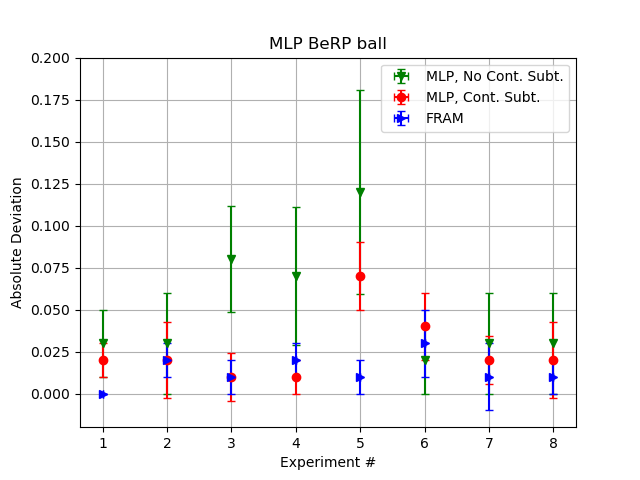}
    \caption{Comparison of absolute deviation from true isotopics ratio for BeRP ball data for various ML algorithms and FRAM (top), and MLP with and without continuum subtraction and FRAM (bottom). Combined statistical and systematic uncertainties are reported at 1 $\sigma$ for all the algorithms except for Decision Trees and Nearest Neighbors. Error bars in FRAM results are the 'sigma' values returned by the FRAM software.}
    \label{fig:BeRP1}
\end{figure}

\begin{figure}
   \centering
    \includegraphics[width=0.45\textwidth]{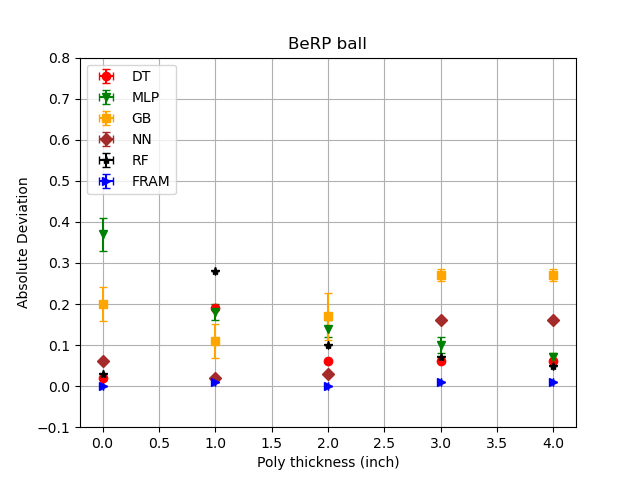}
    \includegraphics[width=0.45\textwidth]{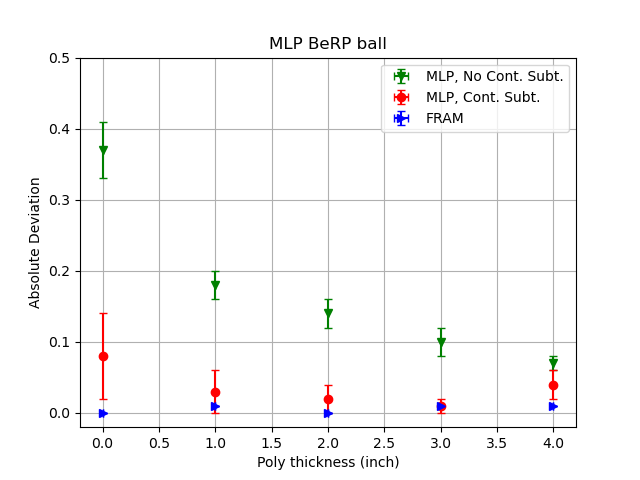}
   \caption{Comparison of absolute deviation from true isotopics ratio for BeRP ball data for various ML algorithms and FRAM (top), and MLP with and without continuum subtraction and FRAM (bottom) as a function of polyethylene (shielding material) thickness in inch. Combined statistical and systematic uncertainties are reported at 1 $\sigma$ for all the algorithms except for Decision Trees and Nearest Neighbors. Error bars in FRAM results are the 'sigma' values returned by the FRAM software.}
    \label{fig:BeRP2}
\end{figure}

The BeRP ball results are presented in Figures~\ref{fig:BeRP1} and ~\ref{fig:BeRP2}. The former figure presents absolute deviation from the true isotopics ratio for experiments with different shielding material combinations and thicknesses, while the latter presents analogous results with a polyethylene shielding of 2.52 cm. As observed previously, the MLP results are comparable to the FRAM results. The bottom figures show improved performance for the MLP method when continuum subtraction is performed in line with the previous observation in the plutonium oxide results.\par
%As previously observed in the plutonium oxide results, the bottom figures show improved performance for the MLP method when continuum subtraction is performed.\par
%%%%%%%%%%%%%%%%%%%%%%%%

%Once the validation on simulations are performed, the ML algorithm is tested on several experimental dataset collected with HPGe detectors. In the first experiment performed by Nuclear Engineering and Nonproliferation group at LANL, fourteen experimental data are taken for 7 Pu and 7 U sources without any shielding materials, and with a 4 mm and 1/8" Pb shielding for Pu and U sources respectively. Data is taken for an integration time of 300s. The experimental data is pre-processed to subtract for the terresterial background contribution as described in Section~\ref{sec:terrbkg}.  %a with a bismuth collimator and a tin filter. 
Upon considering all of the analyzed experimental data, the MLP algorithm performed better than the other ML algorithms evaluated.  The better performance of MLP as compared to the other ML algorithms may be attributed to the large interconnections of the fully connected neural network enabling highly non-linear behavior to be learned more readily. Improvements in the predictions of the MLP was observed for the plutonium data set when continuum subtraction was performed prior to the ML application. Although, the amount of improvement varied experiment to experiment, the largest improvement in absolute deviation was seen for Pu oxide data at a value of roughly 0.3. 

%One possible explanation for the improved performance of the plutonium MLP results with background subtraction With diverse enough training sample, the ML algorithm adequately learns the relationships between the input features in the presence of continuum background; hence, resulting in absolute errors for MLP algorithm that are comparable to the existing FRAM method. 

%However, the opposite trend is observed for the uranium dataset. This dichotomy may be explained by the fact that prominent photo-peaks for plutonium exist at medium energy values, where the photo-peak intensity to the scatter background ratio are much larger than at lower energies, where most of the prominent uranium photo-peaks lie. At the lower energy values, the continuum background is often many orders of magnitude larger than the photo-peaks. Small error in the continuum background estimate has a larger impact in the estimate for the photo-peak count at lower energy values than at higher energy values. Therefore, when the ML is trained and tested on features with continuum subtraction, the error in photo-peak count estimate from the subtraction process may propagate in the way that it results in larger absolute error for uranium enrichment prediction. The MLP algorithm, through its higher inter-connectedness 

\section{Conclusions}
\label{sec:Conclusions}
Several machine learning (ML) based regression algorithms were investigated to perform quantitative determination of uranium and plutonium isotopics using $\mathrm{\gammaup}$-ray spectroscopy data collected with HPGe detectors. The algorithms were trained using GADRAS simulations with different source geometries and thicknesses as well as shielding material types and thicknesses to address the needs of the Emergency Response community. Performance of the algorithms was examined using both simulations as well as experimental datasets incorporating both uranium and plutonium sources in oxide and metal forms.\par
Without time-consuming pre-processing that is often required using conventional methods, all the investigated algorithms were found to offer excellent performance when simulation data was utilized. A slight decrease in performance was observed with increasing complexity, i.e. wider ranges in source thicknesses, shielding conditions, etc. Additional subtraction of the continuum background in the pre-processing stage had a minimum impact in the performance, indicating that ML algorithms were able to adequately learn the feature relationships in the presence of a large continuum background.\par
For the experimental dataset, the results were found to be consistently better using a fully connected neural network (or MLP) algorithm as compared to other algorithms that were investigated. Comparison of these results with results obtained from conventional methods (FRAM software) showed comparable error in the isotopic ratio estimate. Finally, our results demonstrate that with minimum pre-processing, ML algorithms are a good alternative to conventional methods of isotopic determination. The performance of ML algorithms may be enhanced by substantially increasing the training data volume and the physics phase space it covers for improved machine learning interpolation at unknown configurations. 
\bibliography{mybibfile}

\end{document}